\documentclass[useAMS,usenatbib]{mn2e}

\usepackage{graphicx}
\usepackage{amsmath}
\usepackage{txfonts}
\usepackage{natbib}
\usepackage{textcomp}
\usepackage{subfigure}

\title[]{Collisional excitation of water by hydrogen atoms}
\author[F. Daniel et al.]{F. Daniel$^{1}$\thanks{E-mail:
fabien.daniel@obs.ujf-grenoble.fr}, 
A. Faure$^{1}$ ,
P. J. Dagdigian$^{2}$, 
M.-L. Dubernet$^{3,4}$, 
F. Lique$^{5}$,
G. Pineau des for\^ets$^{6,7}$
 \\
$^{1}$ Univ. Grenoble Alpes / CNRS, IPAG, F-38000 Grenoble, France \\
$^{2}$ Department of Chemistry, The Johns Hopkins University, Baltimore, Maryland 21218-2685, USA \\
$^{3}$ Universit\'e Pierre et Marie Curie, LPMAA, UMR CNRS 7092, 75252 Paris, France \\
$^{4}$ Observatoire de Paris, LUTH, UMR CNRS 8102, 92195 Meudon, France \\
$^{5}$ LOMC-UMR 6294, CNRS-Universit\'e du Havre, 25 rue Philippe Lebon, BP 540 76058 Le Havre France \\
$^{6}$ LERMA, UMR 8112, CNRS, Observatoire de Paris, ENS, UPMC, UCP, 61 avenue de l'Observatoire, F-75014 Paris \\
$^{7}$ IAS, UMR 8617, CNRS, B\^atiment 121, Universit\'e Paris Sud 11, 91405, Orsay, France  }
\begin{document}

\date{Accepted XXX. Received XXX; in original form XXX}

\pagerange{\pageref{firstpage}--\pageref{lastpage}} \pubyear{2014}

\maketitle

\label{firstpage}

\begin{abstract}

We present quantum dynamical calculations that describe the rotational
excitation of H$_2$O due to collisions with H atoms. We used a recent,
high accuracy potential energy surface, and solved the collisional
dynamics with the close--coupling formalism, for total energies up to
12 000 cm$^{-1}$. From these calculations, we obtained collisional
rate coefficients for the first 45 energy levels of both ortho-- and
para--H$_2$O and for temperatures in the range T = 5--1500 K. These
rate coefficients are subsequently compared to the values previously
published for the H$_2$O / He and H$_2$O / H$_2$ collisional
systems. It is shown that no simple relation exists between the three
systems and that specific calculations are thus mandatory.

\end{abstract}


\section{Introduction}

In astrophysical studies, water is an important molecule which has
been observed in many different media, ranging from cold prestellar
cores \citep{caselli2010,caselli2012}, to warm intermediate and high
mass star--forming regions \citep{cernicharo1994,cernicharo2006},
circumstellar envelopes \citep[see
  e.g.][]{gonzalez-alfonso1999,decin2010} or extra--galactic sources
\citep{gonzalez-alfonso2004}. In these media, the emission observed
for H$_2$O is often associated with shocked gas
\citep{cernicharo1999,flower2010,neufeld2014}.  An extensive view of
the water component in these regions can be found in the reviews by
\citet{cernicharo2005} and by \citet{vandishoeck2011,vandishoeck2013}.

In order to interpret the H$_2$O line intensities and infer the
physical and chemical properties of the observed regions, the most
reliable methodology relies on radiative transfer calculations. Indeed,
for most of these objects, the water energy levels are often populated
under non--LTE conditions. It is thus necessary to know the H$_2$O
collisional rate coefficients with the relevant collisional partners,
namely H$_2$, He, $e^-$ and H. Extensive
collisional data sets are now available for the three former colliders
\citep{daniel2011,green1993,faure2004,faure2008}. 
The collision between a H$_2$O molecule and H atom has been the subject of many studies \citep[see e.g.][]{jiang2011,fu2013}.
However, most of these works were only focused on reactive collisions and in the formation of the OH and H$_2$ molecules. 
To the best of our knowledge, no quantum state--to--state rate coefficients have been published for the inelastic collisions. 

The molecules for which rate coefficients have been obtained with H as
a collisional partner are limited.  Indeed, in the case of molecules
which have been detected in the interstellar medium, the only
available calculations consider CO
\citep{chu1975,green1976,balakrishnan2002,shepler2007,yang2013},
CO$^{+}$ \citep{andersson2008}, N$_2$ \citep{stoecklin2008},
H$_2$ \citep{forrey1997,flower1998,wrathmall2006,lique2012a} and HD \citep{flower1999,roueff1999}.
To date, no data are available for the inelastic rate coefficients of
the H$_2$O / H collisional system. Therefore, these quantities are
usually inferred by scaling either the H$_2$O / He
\citep[e.g.]{nesterenok2014} or H$_2$O / H$_2$
\citep[e.g.]{flower2010} collisional rate coefficients, when
needed. In the current study, we provide rate coefficients for the
H$_2$O / H system and comment on the possibility of scaling the rates
from other collisional systems to have an estimate of such rates. 
These collision rates are of particular importance in predicting and interpreting H$_2$O line emission in 
shocks propagating in the molecular interstellar medium. Indeed, in J--type shocks with velocities greater 
than 15--20 km/s, the temperature behind the shock front is sufficient to dissociate molecular hydrogen 
and there is a large range of temperature (300--2000K) where H$_2$O and atomic hydrogen coexist and 
where the cooling can be dominated by H$_2$O line emission \citep{flower2003,flower2010}. 
Finally, in such regions, some water lines may exhibit population inversion. In order to interpret the emission from these masers, 
it is necessary to describe accurately the rates at which the upper and lower states are populated, which thus depends 
on the collisional rate coefficients used \citep{daniel2013,hollenbach2013}.

This article is organized as follows. In Sect. \ref{potentiel}, we
describe the potential energy surface used in the current work. In
Sect. \ref{dynamique}, we present the quantum dynamical calculations
and in Sect. \ref{rates}, we discuss the current rate coefficients
with respect to other collisional systems involving the water
molecule.

\section[]{Potential Energy Surface} \label{potentiel}

A high accuracy potential energy surface (PES) for the interaction of
H$_2$O with a hydrogen atom was computed recently by
\citet{dagdigian2013}. The rigid-rotor approximation was employed with
the water geometry kept fixed at its vibrationally averaged
geometry. The reactive channel leading to OH + H$_2$ is thus ignored
and the PES is three-dimensional. The rigid-rotor approximation is
valid at the temperatures investigated here since the activation
energy for the reaction is high ($\sim$ 9300~K) and the rate
coefficient is only $\sim 2\times 10^{-13}$cm$^3$s$^{-1}$ at 1500~K
\citep{baulch1992}. \citet{dagdigian2013} employed restricted coupled
cluster calculations with inclusion of single and double excitations,
augmented by a perturbational estimate of the connected triple
excitations [RCCSD(T)]. A quadruple zeta quality basis set was used,
with the addition of mid--bond functions, and a counterpose correction
was applied to correct for basis set superposition error (BSSE). By
exploiting symmetry, a total of 3800 nuclear geometries only were
computed for atom-molecule separations ranging from 3 to 10 bohr. Full
details about the PES and the H$_2$O--H system can be found in
\citet{dagdigian2013}.

In order to interface the H$_2$O--H potential of \cite{dagdigian2013}
with the MOLSCAT scattering program (see below), it was necessary to
perform a new angular expansion of the PES. Indeed, the coordinates
used by \cite{dagdigian2013} to describe the H$_2$O-H PES (see Fig.~1
of their paper) are different from those required by MOLSCAT for an
atom-asymmetric top system. As a result, the 3800 nuclear geometries
were converted to the MOLSCAT coordinate system where the $z$ axis is
the symmetry axis of water. This conversion corresponds in practice to
a single rotation by an angle of 90$^{\rm o}$ about the $y$ axis,
which is common to both sets of coordinates. The resulting new
spherical coordinates were duplicated in the whole sphere,
i.e. $\theta \in [0,180]^{\rm o}$ and $\phi \in [0,360]^{\rm o}$, both
varied in steps of 10$^{\rm o}$, for a total of 12~620 nuclear
geometries (631 per intermolecular separation).

The H$_2$O--H PES expressed in the MOLSCAT coordinate system was
expanded in spherical harmonics, $Y_{\lambda, \mu}(\theta,\phi)$, as
in \citet{dagdigian2013} (see their Eq.~13), using a linear
least-square fit procedure\footnote{We note that in the MOLSCAT
  coordinate system, the $C_{2v}$ symmetry of H$_2$O requires that
  $\mu$ is even, while in the original coordinate system, symmetry
  restricts the expansion to terms with $\lambda + \mu$ even.}. As in
\citet{dagdigian2013}, all terms with $\lambda\leq 10$ and $\mu \leq
8$ were included in the expansion, resulting in a total of 35 angular
terms. The root mean square residual was found to be lower than
1~cm$^{-1}$ for intermolecular separations $R$ larger than 4~bohr. As
an illustrative example, the global minimum of the PES was found by
\citet{dagdigian2013} to have an energy of -61.0~cm$^{-1}$, at a
geometry of $R$=6.5~bohr, $\theta=120^{\rm o}$, $\phi=0^{\rm o}$ (in
MOLSCAT coordinates). 
%
%
Using our fit, we obtained -61.3~cm$^{-1}$, in
excellent agreement. This latter value can also be compared with the
global minima of the PES for the similar systems H$_2$O-He
(-34.9~cm$^{-1}$ at $R$=5.9~bohr, $\theta=75^{\rm o}$, $\phi=0^{\rm o}$ \citet{patkowski2002}) and H$_2$O-H$_2$
(-235.1~cm$^{-1}$ at $R$=5.8~bohr, $\theta=0^{\rm o}$, $\phi=0^{\rm o}$, \citet{faure2005,valiron2008}). The interaction of
water with hydrogen atoms is thus very different from the interactions
with He and H$_2$. We can therefore expect significant differences in
the corresponding rotational rate coefficients.

\section[]{Collisional dynamics} \label{dynamique}

In order to solve the collisional dynamics, we used the 
MOLSCAT\footnote{J. M. Hutson and S. Green, MOLSCAT computer code,
  version 14 (1994), distributed by Collaborative Computational
  Project No. 6 of the Engineering and Physical Sciences Research
  Council (UK).} code. Benchmark calculations were also 
performed with the HIBRIDON\footnote{HIBRIDON is a package of programs for the time-independent
quantum treatment of inelastic collisions and photodissociation written
by M. H. Alexander, D. E. Manolopoulos, H.-J. Werner, B. Follmeg,
Q. Ma, and P. J. Dagdigian, with contributions by P. F. Vohralik, D.
Lemoine, G. Corey, R. Gordon, B. Johnson, T. Orlikowski, A. Berning,
A. Degli-Esposti, C. Rist, B. Pouilly, G. van der Sanden, M. Yang, F.
de Weerd, S. Gregurick, J. Klos and F. Lique. More information and/or
a copy of the code can be obtained from the website http://www2.
chem.umd.edu/groups/alexander/hibridon/hib43.} code using the original fit of the H$_2$O--H PES 
by \citet{dagdigian2013}. We tested both codes at a few total energies and the cross sections were found to 
be essentially similar, within 5\%.
We performed the calculations in order to provide
rate coefficients for the first 45 energy levels of the ortho-- and
para--H$_2$O symmetries, i.e. up to $J_{K_a,K_c} = 7_{7,0}$ (E $\sim$
1395 cm$^{-1}$) for o--H$_2$O and up to $J_{K_a,K_c} = 7_{7,1}$ (E
$\sim$ 1395 cm$^{-1}$) for p--H$_2$O.  Calculations have been
performed up to a total energy of 12 000 cm$^{-1}$ and we have used
the close--coupling formalism over the whole energy range.  This
enables to provide converged rate coefficients for the range of
temperature $T = 5-1500$K.  The parameters of the calculations,
i.e. the number of H$_2$O energy levels, the integration step and the
step between two consecutive energies were determined in order to
ensure an accuracy better than 5\% for the rate coefficients. These
parameters are given in Tables~\ref{table:param_conv} and
\ref{table:E_step}. Additionally, we included a cut in energy for the
H$_2$O energy levels, set to $E_{max}$ = 3000 cm$^{-1}$ below total
energy of 5000 cm$^{-1}$ and $E_{max}$ = 4000 cm$^{-1}$ above this
threshold.

The water energy levels are described using the effective Hamiltonian
of \citet{kyro1981}, as previously done in our quantum calculations
that dealt with the H$_2$O / H$_2$ system
\citep{dubernet2002,grosjean2003,dubernet2006,dubernet2009,daniel2010,daniel2011}.
We used the hybrid modified log-derivative Airy propagator of
\citet{alexander1987}, the change of propagator being set at 20
$a_0$. The reduced mass of the collisional system is $\mu$ =
0.954418234 amu.

\begin{table}
\caption{Parameters that govern the convergence of the MOLSCAT calculations: \texttt{J1max} which is the highest value
of the rotational quantum number and the parameter \texttt{STEPS} which is inversely proportional to the 
step of integration. These parameters are given for the two water symmetries as a function of the total energy.}
\begin{center}
\begin{tabular}{ccccc}
\hline
 & \multicolumn{2}{c}{p--H$_2$O} & \multicolumn{2}{c}{o--H$_2$O} \\
   Energy range (cm$^{-1}$) & \texttt{J1max} & \texttt{STEPS} &  \texttt{J1max} & \texttt{STEPS} \\ \hline
          $<$ 42       &   4  &  20   &   4  &  50  \\
         42--80         &   4  &  10   &   4  &  20  \\          
         80--230       &   6  &  10   &   5  &  10  \\
       230--310       &   6  &  10   &   6  &  10  \\
       310--410       &   7  &  10   &   7  &  10  \\
       410--500       &   8  &  10   &   8  &  10  \\
       500--600       &   9  &  10   &   9  &  10  \\
       600--750       &  10  &  10  &  10  &  10  \\
       750--1000     &  11  &  10  &  11  &  10  \\
     1000--1250     &  12  &  10  &  12  &  10   \\
     1250--1500     &  13  &  10  &  13  &  10  \\
     1500--2000     &  14  &  10  &  14  &  10  \\
     2000--3000     &  15  &  10  &  15  &  10  \\
     3000--5000     &  16  &  10  &  16  &  10   \\
     5000--12000   &  18  &  10  &  18  &  10   \\   \hline
\end{tabular}
\end{center}
\label{table:param_conv}
\end{table}%

\begin{table}
\caption{Step between the consecutive total energies used to characterize the cross sections.}
\begin{center}
\begin{tabular}{cc}
\hline
 Energy range (cm$^{-1}$) & step in energy (cm$^{-1}$) \\ \hline
         $<$ 1250  & 0.1  \\
  1250 $-$ 2000 & 0.5  \\ 
  2000 $-$ 2500 & 5.0  \\ 
  2500 $-$ 3000 & 10.0 \\
  3000 $-$ 12000 & 50.0 \\ \hline
\end{tabular}
\end{center}
\label{table:E_step}
\end{table}%

\section{Rate coefficients} \label{rates}

The collisional de-excitation rate coefficients are calculated by averaging 
the cross sections with a Maxwell--Boltzmann distribution that describes the 
distribution of velocity of the molecules in the gas \citep[see e.g. eq. (2) in][]{dubernet2006}
\begin{eqnarray}
R_{\beta,\beta'}(T) = \left( \frac{8}{\mu \pi} \right)^{\frac{1}{2}} 
\frac{1}{ \left ( k_B \, T \right )^{\frac{3}{2}}}
\int_0^{\infty} \sigma_{\beta,\beta'}(E) \, E \, e^{- E / k_B \, T } \, \textrm{d}E
\end{eqnarray}
where $\beta$ and $\beta'$ are a set of quantum numbers that describe
the initial and final states of water, $k_B$ is the boltzmann
constant, $\mu$ is the reduced mass of the colliding system and $E$ is
the kinetic energy.
In Table \ref{table:rates}, we give the de--excitation rate
coefficients for levels up to $J_{Ka,Kc} = 3_{3,0}$, and for
temperatures in the range T = 20--1000 K.  The whole set of rate
coefficients, with higher temperatures and with a more extended set of
molecular levels will be made available through the LAMDA
\citep{schoier2005} and BASECOL \citep{dubernet2013} databases.

\begin{figure}
\begin{center}
\includegraphics[angle=0,scale=0.45]{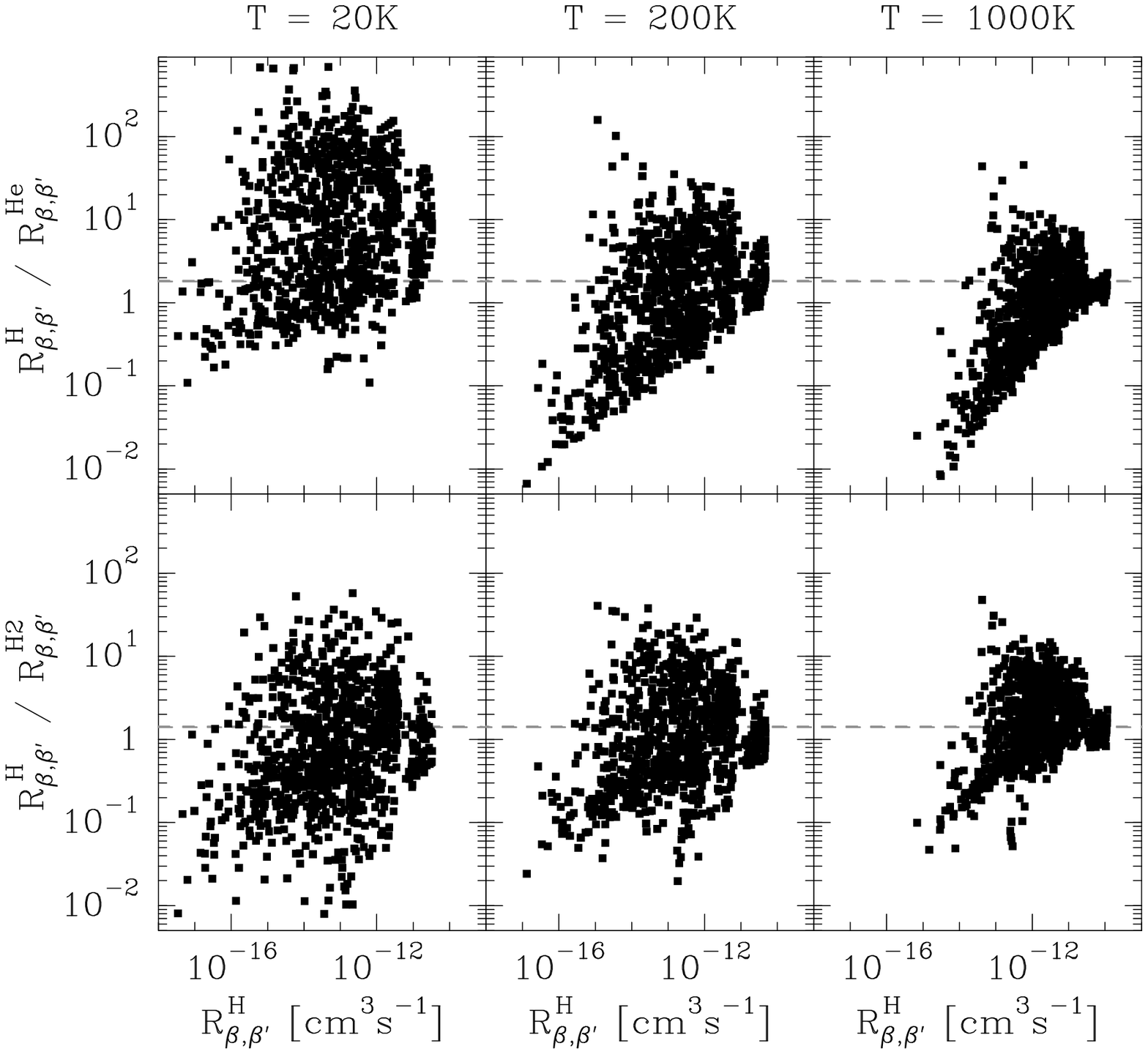}
\caption{Ratios of the He \citep[$R_{\beta,\beta'}^{\textrm{He}}$ from][]{green1993} 
and H$_2$ \citep[$R_{\beta,\beta'}^{\textrm{H2}}$ from][]{dubernet2009} collisional rate coefficients with the o--H$_2$O rates calculated with H ($R_{\beta,\beta'}^{\textrm{H}}$) as a collider. These values are given as a function of the magnitude of $R_{\beta,\beta'}^{\textrm{H}}$ and for temperatures of 20K (first column) 200K (second column) and 1000K (third column).}
\label{fig:ratios}
\end{center}
\end{figure}

In astrophysical applications, it is rather common to need rate
coefficients which are not available. Therefore, it is quite usual to
infer the rate coefficients of a colliding system from the values
calculated for closely related system. The methodology which is
generally used, even if its theoretical basis are questionable
\citep{walker2014}, consist in assuming that the cross sections
$\sigma_{\beta,\beta'}(E)$ which appear in eq. (1), are similar for
both systems.  The rate coefficients are then derived by correcting
for the change in reduced mass, which lead to the scaling
relationships: $R^2_{\beta,\beta'}(T) = \sqrt{\mu_1/\mu_2} \times
R^1_{\beta,\beta'}(T)$.  As an example, in the current case, we could
apply this methodology to infer the H$_2$O / H rate coefficients from
either the H$_2$O / He or H$_2$O / p--H$_2$ rate coefficients. This
would lead to rate coefficients such that $R^\textrm{H}_{\beta,\beta'}
\sim 1.8 \times R^{\textrm{He}}_{\beta,\beta'}$ or
$R^\textrm{H}_{\beta,\beta'} \sim 1.4 \times
R^{\textrm{H}_2}_{\beta,\beta'}$. In table \ref{table:rates}, we
compare the current H$_2$O / H rate coefficients with the rate
coefficients of the H$_2$O / He \citep{yang2013} and H$_2$O / p--H$_2$
\citep{dubernet2009} systems, for o--H$_2$O energy levels up to
$J_{K_a,K_c} = 3_{3,0}$ (E $\sim$ 285K).
In Fig. \ref{fig:ratios}, we give the ratios of the current o--H$_2$O
/ H rate coefficients with the He rate coefficients of
\citet{green1993} and with the H$_2$ rates of \citet{dubernet2009},
for the first 45 o--H$_2$O energy levels. In the case of He, we used
the rate coefficients from \citet{green1993} since the most recent
calculations by \citet{yang2013} only consider the first 10 water
energy levels. Additionally, as pointed out in the latter study, the
impact of the new calculation is modest at high temperatures, the
differences being lower than 30\% above 200K. At lower temperatures,
however, differences of up to a factor 3 can be found between the two
sets.
Considering the values taken by the ratios, it is obvious that no
simple scaling relationship relate the various collisional rate
coefficient sets, with differences that span various orders of
magnitude. However, we note that the scatter of the ratios tends to
decrease when increasing temperature. Moreover, for the largest rates
at high temperature, the rate coefficients for the various colliders
become similar, within a factor 2. This was expected since at high
collision energy the scattering process becomes dominated by
kinematics rather than specific features of the PES. Such conclusions
were already reached for other collisional systems \citep[see
  e.g.][for the HD molecule]{roueff1999}. In summary, a dedicated
calculation is a pre--requisite to accurately describe the collision
with either H$_2$, He or H, especially at temperatures below
$\sim$1000~K.

\section{Conclusions}

We described quantum dynamical calculations performed at the
close--coupling level of theory for the H$_2$O / H collisional
system. As a result, we give collisional rate coefficients for the
first 45 energy levels of both ortho-- and para--H$_2$O and for
temperatures in the range T = 5--1500 K. These calculations complete
the sets already calculated for the water molecule, for which specific
calculations are now available for all the colliders relevant to
studies dealing with the interstellar medium, i.e. H$_2$, He, H and
$e^-$.  In particular, we examined the possibility of emulating the
H$_2$O / H rate coefficients by a simple scaling of either the H$_2$O
/ He or H$_2$O / H$_2$ sets and found that no simple relation enable
to relate a set to another.

\begin{table*}
\caption{H$_2$O / H  de--excitation rate coefficients (R$^\textrm{H}_{\beta,\beta'}$ in cm$^3$ s$^{-1}$, \textbf{where the notation $x(y)$ stands for $x \, 10^y$ }) as a function of temperature and for H$_2$O levels up to $J_{Ka,Kc} = 3_{3,0}$.
For each temperature, we give the ratio of the current rate coefficients with the He rate coefficients of \citet{yang2013} (second column), and with 
the para--H$_2$ rate coefficients of \citet{dubernet2009} (third column). }
\begin{center}
\begin{tabular}{ccccccccccccc}
\hline \hline
 Transition & \multicolumn{3}{c}{20K} & \multicolumn{3}{c}{100K} & \multicolumn{3}{c}{500K} & \multicolumn{3}{c}{1000K} \\
 & R$^\textrm{H}_{\beta,\beta'}$  & H / He & H / H$_2$ 
 & R$^\textrm{H}_{\beta,\beta'}$  & H / He & H / H$_2$ 
 & R$^\textrm{H}_{\beta,\beta'}$  & H / He & H / H$_2$ 
 & R$^\textrm{H}_{\beta,\beta'}$  & H / He & H / H$_2$ \\ \hline
$1_{ 1,  0 }  \to 1_{ 0,  1 } $  &    9.54(-12) &  0.78 &  0.28 &    1.88(-11) &  0.62 &  0.46 &    6.84(-11) &  0.85 &  0.70 &    1.06(-10) &  1.07 &  0.83 \vspace{0.2cm} \\
$2_{ 1,  2 }  \to 1_{ 0,  1 } $  &    1.24(-11) &  0.68 &  0.37 &    1.71(-11) &  0.62 &  0.46 &    5.24(-11) &  0.89 &  0.75 &    8.09(-11) &  1.11 &  0.94 \\
$2_{ 1,  2 }  \to 1_{ 1,  0 } $  &    6.34(-13) &  0.11 &  0.04 &    9.66(-13) &  0.14 &  0.08 &    3.54(-12) &  0.40 &  0.42 &    6.39(-12) &  0.71 &  0.86 \vspace{0.2cm}\\
$2_{ 2,  1 }  \to 1_{ 0,  1 } $  &    8.60(-13) &  1.22 &  0.26 &    1.86(-12) &  0.89 &  0.40 &    9.00(-12) &  0.73 &  0.80 &    1.50(-11) &  0.70 &  0.86 \\
$2_{ 2,  1 }  \to 1_{ 1,  0 } $  &    1.80(-11) &  2.07 &  0.54 &    2.00(-11) &  1.16 &  0.56 &    4.90(-11) &  1.09 &  0.79 &    7.36(-11) &  1.27 &  0.96 \\
$2_{ 2,  1 }  \to 2_{ 1,  2 } $  &    6.32(-12) &  0.57 &  0.34 &    9.07(-12) &  0.53 &  0.47 &    2.62(-11) &  0.78 &  0.68 &    3.91(-11) &  0.99 &  0.80 \vspace{0.2cm}\\
$3_{ 0,  3 }  \to 1_{ 0,  1 } $  &    7.51(-13) &  0.49 &  0.10 &    1.48(-12) &  0.42 &  0.19 &    6.91(-12) &  0.57 &  0.74 &    1.23(-11) &  0.70 &  1.11 \\
$3_{ 0,  3 }  \to 1_{ 1,  0 } $  &    2.61(-12) &  0.51 &  0.43 &    3.60(-12) &  0.49 &  0.60 &    1.09(-11) &  0.90 &  1.15 &    1.78(-11) &  1.19 &  1.69 \\
$3_{ 0,  3 }  \to 2_{ 1,  2 } $  &    9.32(-12) &  0.68 &  0.27 &    1.28(-11) &  0.61 &  0.40 &    3.68(-11) &  0.82 &  0.65 &    5.56(-11) &  1.02 &  0.81 \\
$3_{ 0,  3 }  \to 2_{ 2,  1 } $  &    1.18(-12) &  2.16 &  0.27 &    7.44(-13) &  0.36 &  0.32 &    3.19(-12) &  0.30 &  0.79 &    7.83(-12) &  0.52 &  1.10 \vspace{0.2cm}\\
$3_{ 1,  2 }  \to 1_{ 0,  1 } $  &    5.32(-13) &  1.68 &  0.66 &    5.68(-13) &  1.01 &  0.73 &    1.34(-12) &  1.18 &  0.96 &    2.11(-12) &  1.36 &  1.18 \\
$3_{ 1,  2 }  \to 1_{ 1,  0 } $  &    1.02(-12) &  1.16 &  0.28 &    1.85(-12) &  0.87 &  0.40 &    9.47(-12) &  1.07 &  1.36 &    1.74(-11) &  1.23 &  2.10 \\
$3_{ 1,  2 }  \to 2_{ 1,  2 } $  &    1.41(-12) &  0.33 &  0.08 &    2.97(-12) &  0.41 &  0.18 &    1.56(-11) &  0.63 &  0.82 &    2.77(-11) &  0.73 &  1.22 \\
$3_{ 1,  2 }  \to 2_{ 2,  1 } $  &    3.00(-12) &  0.69 &  0.18 &    3.76(-12) &  0.50 &  0.31 &    1.47(-11) &  0.94 &  0.91 &    2.63(-11) &  1.30 &  1.44 \\
$3_{ 1,  2 }  \to 3_{ 0,  3 } $  &    7.80(-12) &  0.71 &  0.37 &    1.42(-11) &  0.71 &  0.53 &    5.12(-11) &  0.95 &  0.83 &    8.15(-11) &  1.17 &  1.00 \vspace{0.2cm}\\
$3_{ 2,  1 }  \to 1_{ 0,  1 } $  &    1.69(-12) &  2.57 &  2.18 &    2.34(-12) &  1.24 &  1.91 &    9.59(-12) &  0.93 &  1.83 &    1.72(-11) &  1.01 &  1.84 \\
$3_{ 2,  1 }  \to 1_{ 1,  0 } $  &    2.39(-13) &  1.49 &  0.57 &    2.80(-13) &  0.71 &  0.48 &    4.63(-13) &  0.40 &  0.38 &    6.39(-13) &  0.42 &  0.38 \\
$3_{ 2,  1 }  \to 2_{ 1,  2 } $  &    1.17(-11) &  3.14 &  0.84 &    1.28(-11) &  1.58 &  0.80 &    3.05(-11) &  1.24 &  1.06 &    4.76(-11) &  1.36 &  1.30 \\
$3_{ 2,  1 }  \to 2_{ 2,  1 } $  &    1.32(-12) &  0.28 &  0.07 &    2.77(-12) &  0.36 &  0.13 &    1.48(-11) &  0.78 &  0.82 &    2.76(-11) &  1.01 &  1.57 \\
$3_{ 2,  1 }  \to 3_{ 0,  3 } $  &    6.24(-13) &  0.76 &  0.21 &    7.48(-13) &  0.53 &  0.26 &    2.76(-12) &  0.41 &  0.50 &    4.95(-12) &  0.43 &  0.55 \\
$3_{ 2,  1 }  \to 3_{ 1,  2 } $  &    7.88(-12) &  0.60 &  0.34 &    1.49(-11) &  0.67 &  0.49 &    5.17(-11) &  0.91 &  0.75 &    8.08(-11) &  1.11 &  0.90 \vspace{0.2cm}\\
$4_{ 1,  4 }  \to 1_{ 0,  1 } $  &    2.92(-12) &  1.93 &  0.59 &    3.14(-12) &  1.08 &  0.64 &    7.97(-12) &  1.22 &  1.21 &    1.31(-11) &  1.43 &  1.84 \\
$4_{ 1,  4 }  \to 1_{ 1,  0 } $  &    2.56(-13) &  0.22 &  0.29 &    4.30(-13) &  0.16 &  0.36 &    2.48(-12) &  0.22 &  0.53 &    5.24(-12) &  0.32 &  0.62 \\
$4_{ 1,  4 }  \to 2_{ 1,  2 } $  &    1.00(-12) &  1.99 &  0.33 &    1.43(-12) &  1.17 &  0.35 &    5.55(-12) &  1.43 &  1.05 &    1.05(-11) &  1.81 &  1.84 \\
$4_{ 1,  4 }  \to 2_{ 2,  1 } $  &    9.88(-13) &  0.50 &  0.52 &    1.39(-12) &  0.43 &  0.61 &    3.92(-12) &  0.49 &  0.93 &    6.30(-12) &  0.56 &  1.11 \\
$4_{ 1,  4 }  \to 3_{ 0,  3 } $  &    2.00(-11) &  1.32 &  0.40 &    2.24(-11) &  0.91 &  0.46 &    5.41(-11) &  0.94 &  0.72 &    8.01(-11) &  1.10 &  0.88 \\
$4_{ 1,  4 }  \to 3_{ 1,  2 } $  &    1.56(-13) &  0.16 &  0.08 &    4.46(-13) &  0.29 &  0.18 &    9.40(-13) &  0.20 &  0.34 &    2.01(-12) &  0.27 &  0.46 \\
$4_{ 1,  4 }  \to 3_{ 2,  1 } $  &    7.27(-13) &  0.78 &  0.26 &    1.23(-12) &  0.43 &  0.31 &    4.88(-12) &  0.65 &  0.61 &    8.47(-12) &  0.92 &  0.90 \vspace{0.2cm}\\
$3_{ 3,  0 }  \to 1_{ 0,  1 } $  &    3.03(-13) & 61.35 &  5.14 &    4.20(-13) & 17.93 &  4.59 &    2.15(-12) &  4.01 &  6.86 &    4.86(-12) &  3.80 &  5.48 \\
$3_{ 3,  0 }  \to 1_{ 1,  0 } $  &    2.08(-12) &  7.12 &  1.82 &    2.64(-12) &  2.59 &  1.91 &    8.23(-12) &  1.16 &  1.67 &    1.34(-11) &  1.04 &  1.49 \\
$3_{ 3,  0 }  \to 2_{ 1,  2 } $  &    8.40(-13) &  2.66 &  0.86 &    1.06(-12) &  1.13 &  0.85 &    3.44(-12) &  0.67 &  0.90 &    5.61(-12) &  0.66 &  0.89 \\
$3_{ 3,  0 }  \to 2_{ 2,  1 } $  &    2.71(-11) &  4.51 &  0.80 &    2.92(-11) &  2.04 &  0.76 &    5.89(-11) &  1.28 &  0.88 &    8.29(-11) &  1.35 &  1.01 \\
$3_{ 3,  0 }  \to 3_{ 0,  3 } $  &    4.90(-13) &  1.08 &  0.46 &    4.70(-13) &  0.47 &  0.39 &    1.29(-12) &  0.24 &  0.54 &    2.97(-12) &  0.31 &  0.79 \\
$3_{ 3,  0 }  \to 3_{ 1,  2 } $  &    5.68(-13) &  0.66 &  0.31 &    6.31(-13) &  0.34 &  0.41 &    4.16(-12) &  0.53 &  1.10 &    9.61(-12) &  0.82 &  1.49 \\
$3_{ 3,  0 }  \to 3_{ 2,  1 } $  &    8.63(-12) &  0.73 &  0.38 &    1.07(-11) &  0.63 &  0.44 &    3.05(-11) &  0.94 &  0.78 &    4.82(-11) &  1.25 &  1.04 \\
$3_{ 3,  0 }  \to 4_{ 1,  4 } $  &    5.47(-13) &  0.76 &  0.41 &    3.98(-13) &  0.21 &  0.40 &    2.23(-12) &  0.23 &  0.74 &    5.30(-12) &  0.36 &  0.95 \vspace{0.2cm}\\
\hline
\end{tabular}
\end{center}
\label{table:rates}
\end{table*}%

\section*{Acknowledgments}
Most of the computations presented in this paper were
performed using the CIMENT infrastructure
(https://ciment.ujf-grenoble.fr), which is supported by the
Rh\^one-Alpes region (GRANT CPER07\_13 CIRA: http://www.ci-ra.org).
This work has been supported by the Agence Nationale de la Recherche
(ANR-HYDRIDES), contract ANR-12-BS05-0011-01 and by the CNRS national
program ``Physico-Chimie du Milieu Interstellaire''.

\bsp

\label{lastpage}

\end{document}